\documentclass[journal,12pt,onecolumn,draftclsnofoot,]{IEEEtran}
\IEEEoverridecommandlockouts
\usepackage{mathtools}
\usepackage{bm}
\usepackage{graphicx}
\usepackage{booktabs}
\usepackage{algorithm}%
\usepackage{algpseudocode}%
\usepackage{cite}
\usepackage{url}
\usepackage{microtype}
\usepackage{siunitx}
\usepackage{amsmath,amsfonts,amssymb,amsthm, bm}
\newcommand{\RR}{\mathbb{R}}
\newcommand{\1}{\mathbb{I}}
\newcommand{\rev}[1]{\textcolor{black}{#1}}

\usepackage{scalerel}

\usepackage{pgfplots}
\pgfplotsset{compat=1.18}
\usepackage{tikz}
\usetikzlibrary{arrows.meta,positioning,fit,calc,shapes.geometric,shapes.multipart}
\usetikzlibrary{positioning,calc,arrows.meta}
\usetikzlibrary{positioning,arrows.meta,fit,backgrounds,decorations.pathreplacing}
\tikzset{
  arrow/.style={-Latex, line width=0.8pt},
  block/.style={draw, rounded corners=2pt, align=center, minimum height=6mm, inner sep=2pt, font=\small},
  op/.style={block, fill=gray!10},
  var/.style={block, fill=black!5},
  gate/.style={block, fill=orange!12},
  delay/.style={block, fill=yellow!15},
  legendbox/.style={draw, rounded corners=2pt, inner sep=2pt, font=\scriptsize, fill=white},
  lbl/.style={font=\scriptsize, inner sep=1pt}
}
\usetikzlibrary{svg.path}

\definecolor{orcidlogocol}{HTML}{A6CE39}
\tikzset{
  orcidlogo/.pic={
    \fill[orcidlogocol] svg{M256,128c0,70.7-57.3,128-128,128C57.3,256,0,198.7,0,128C0,57.3,57.3,0,128,0C198.7,0,256,57.3,256,128z};
    \fill[white] svg{M86.3,186.2H70.9V79.1h15.4v48.4V186.2z}
                 svg{M108.9,79.1h41.6c39.6,0,57,28.3,57,53.6c0,27.5-21.5,53.6-56.8,53.6h-41.8V79.1z M124.3,172.4h24.5c34.9,0,42.9-26.5,42.9-39.7c0-21.5-13.7-39.7-43.7-39.7h-23.7V172.4z}
                 svg{M88.7,56.8c0,5.5-4.5,10.1-10.1,10.1c-5.6,0-10.1-4.6-10.1-10.1c0-5.6,4.5-10.1,10.1-10.1C84.2,46.7,88.7,51.3,88.7,56.8z};
  }
}

\newcommand\orcidicon[1]{\href{https://orcid.org/#1}{\mbox{\scalerel*{
\begin{tikzpicture}[yscale=-1,transform shape]
\pic{orcidlogo};
\end{tikzpicture}
}{|}}}}

\usepackage{hyperref}
\expandafter\def\expandafter\normalsize\expandafter{%
    \normalsize%
    \setlength\abovedisplayskip{0pt}%
    \setlength\belowdisplayskip{0pt}%
    \setlength\abovedisplayshortskip{-2.5pt}%
    \setlength\belowdisplayshortskip{0pt}%
}

\begin{document}

\title{NOS-Gate: Queue-Aware Streaming IDS for Consumer Gateways under Timing-Controlled Evasion}

\author{ Muhammad~Bilal~\orcidicon{0000-0003-4221-0877},~\IEEEmembership{Senior~Member,~IEEE,} Omer Tariq~\orcidicon{0000-0002-1771-6166}~Hasan Ahmed
\IEEEcompsocitemizethanks{%

\IEEEcompsocthanksitem Muhammad Bilal and Hasan Ahmed are with the School of Computing and Communications, Lancaster University, LA1 4WA, Lancaster, U.K. E-mail: m.bilal@ieee.org, h.ahmed@lancaster.ac.uk.

Omer Tariq is with the School of Computing, Korea Advanced Institute of Science and Technology, Daejeon, 34141, South Korea. E-mail: omertariq@kaist.ac.kr

Published in: IEEE Transactions on Consumer Electronics. 
DOI:10.1109/TCE.2026.3682516

}}

\maketitle

\begin{abstract}
Timing and burst patterns can leak through encryption, and an adaptive adversary can exploit them. This undermines metadata-only detection in a stand-alone  consumer gateway. Therefore, consumer gateways need streaming intrusion detection on encrypted traffic using metadata only, under tight CPU and latency budgets. We present a streaming IDS for stand-alone gateways that instantiates a lightweight two-state unit derived from Network-Optimised Spiking (NOS) dynamics \cite{bilal2025nos} per flow, named  \emph{NOS-Gate}. NOS-Gate scores fixed-length windows of metadata features and, under a $K$-of-$M$ persistence rule, triggers a reversible mitigation that temporarily reduces the flow's weight under weighted fair queueing (WFQ). We evaluate NOS-Gate under timing-controlled evasion using an executable \emph{worlds} benchmark that specifies benign device processes, auditable attacker budgets, contention structure, and packet-level WFQ replay to quantify queue impact. All methods are calibrated label-free via burn-in quantile thresholding. Across multiple reproducible worlds and malicious episodes, at an achieved $0.1\%$ false-positive operating point, NOS-Gate attains 0.952 incident recall versus 0.857 for the best baseline in these runs. Under gating, it reduces p99.9 queueing delay and p99.9 collateral delay with a mean scoring cost of $\approx 2.09\,\mu\mathrm{s}$ per flow-window on CPU. 
\end{abstract}

\section{Introduction}

Home gateways and consumer hubs sit between consumer devices and encrypted services (web and cloud). Encryption limits payload inspection at the edge, but it does not remove traffic side channels such as timing, packet sizes, and burst structure \cite{conti16android,panchenko16wfscale}. These signals have been used for device identification and security monitoring \cite{sivanathan19iotclass,miettinen17iotsentinel,hafeez20iotkeeper,nguyen19diot}, but the gateway setting imposes two constraints that shape what is practical. First, detection must run continuously with low latency and modest CPU, which makes large models and heavy training difficult to justify \cite{chandola09anomalysurvey,mirsky18kitsune}. Second, an attacker can often change packet timing without changing endpoints, ports, or encrypted content.

Timing control is therefore a realistic evasion surface. Prior work shows that timing and burst patterns can still leak structure under encryption \cite{panchenko16wfscale}. The same idea can be used offensively: an attacker can reshape inter-arrival times and bursts to resemble benign device rhythms while still pursuing a hidden attack objective \cite{wright09trafficmorphing,cai14wfdefenses}. This raises an operational IDS question that is not captured well by offline accuracy alone: can a gateway detect quickly enough to act, at a low false-alarm rate, using only metadata, against timing-shaped adversaries?

We address this question with \emph{NOS-Gate}, a streaming IDS for stand-alone gateways derived from Network-Optimised Spiking (NOS) dynamics \cite{bilal2025nos,bilalxunos6g}. NOS-Gate instantiates a lightweight two-state dynamical unit per flow. An excitability-like state accumulates evidence from normalised metadata deviations, while a recovery-like state provides short-term memory and suppresses repeated triggering. The detector emits an event-like surrogate and applies a persistence rule to stabilise decisions. This design is deliberately simple: compact state updates fit gateway budgets while still expressing persistence and hysteresis, which are hard to add to a simple window-based score with no memory. \cite{tavanaei19snnsurvey,markovic20neuromorphic}. 

Crucially, we connect detection to a gateway-safe action. When alarms persist, NOS-Gate temporarily reduces the weight of the flagged flow under weighted fair queueing (WFQ). This aims to reduce queue-tail harm while limiting collateral impact on benign traffic \cite{demers89fq,parekh93gps1,parekh94gps2,nichols12codel}. Figure~\ref{fig:system_overview} summarises the online loop (windowing, scoring, persistence, and WFQ gating) and the offline evaluation harness used to measure queue impact.

Recent encrypted-traffic security systems have improved robustness and deployment realism, but they usually make different assumptions than a consumer gateway IDS. HyperVision detects unknown encrypted malicious traffic using flow-interaction graphs \cite{fu23hypervision}. RAPIER studies detection under low-quality training data \cite{qing24rapier}. Wedjat targets evasive behaviour using real-time causal analysis \cite{gao25wedjat}. ET-SSL explores self-supervised contrastive learning for encrypted-traffic anomaly detection \cite{sattar25etssl}. These works reinforce that encrypted side channels remain useful, but they typically do not evaluate the full gateway action loop (detection $\rightarrow$ mitigation $\rightarrow$ queue impact) under an explicit timing-control budget, and they do not standardise a stand-alone label-free calibration protocol. NOS-Gate is built around those gateway requirements. \rev{Table~\ref{tab:nos_to_nosgate} provides an explicit mapping from NOS symbols to their IDS interpretation in NOS-Gate, so readers can audit what is reused unchanged and what is reinterpreted for detection and action.}

A second contribution is evaluation methodology. Public IoT corpora and botnet traces are valuable, but they often provide limited coverage of timing-controlled evasion and rarely offer an auditable link from attacker constraints to queue-level harm at the gateway \cite{meidan18nbaiot,sicari15iotsec,tahaei20iottraffic}. We therefore introduce an executable \emph{worlds} benchmark that specifies benign device processes and contention structure, constrains adversaries by auditable budgets, and uses packet-level WFQ replay to quantify user-visible impact. We evaluate all detectors under the same stand-alone, label-free protocol: an assumed-benign burn-in segment, per-flow thresholds set by high quantiles of burn-in scores, and labels used only for reporting on held-out time \cite{chandola09anomalysurvey,buczak16survey,mirsky18kitsune}.

\rev{Encrypted-traffic IDS work often improves detection performance under a dataset protocol, but at least one gateway-facing element is commonly left implicit: (i) a stand-alone, label-free protocol that fixes an explicit false-alarm budget without supervised tuning, (ii) an explicit timing-control threat model with auditable operational budgets, and (iii) an action-impact evaluation that measures whether mitigation improves user-visible queue outcomes rather than only producing alarms. This paper adds all three in a single executable loop: NOS-Gate as a compact two-state streaming detector calibrated without labels, \emph{worlds} as a reproducible benchmark with auditable timing and contention budgets (including feasibility reporting), and packet-level WFQ replay to quantify queue-tail and collateral impact under reversible gating.}

We make three contributions that together support a practical, auditable gateway IDS under timing-controlled evasion:
\begin{itemize}
\item We propose a two-state detector \emph{NOS-Gate}, a streaming, metadata-only gateway IDS with a direct action hook: a per-flow two-state detector with persistence-based alerting that drives WFQ deprioritisation.

\item We introduce a reproducible \emph{worlds} benchmark with published contention structure, auditable attacker budgets, and packet-level WFQ replay, enabling queue-tail and collateral evaluation under timing-shaped attacks.

\item We standardise a stand-alone operating protocol at explicit false-alarm budgets using burn-in quantile thresholding and persistence, and we report achieved false-positive rates, incident recall, and time-to-detect under the same calibration rule.
\end{itemize}

\rev{We do not claim payload inspection, DPI deployment in ISP-grade firmware, or that our synthetic worlds reproduce any single public dataset. Instead, under the published feature contract and attacker budgets, NOS-Gate provides a stronger detection-and-action trade-off than the evaluated label-free baselines: higher incident detection at the same achieved false-alarm budget, and improved queue-tail outcomes when WFQ gating is enabled.} This work do not claim payload inspection, DPI deployment in ISP-grade firmware, or that synthetic worlds reproduce any single public dataset. We present the benchmark as a falsifiable protocol: new worlds, budgets, and contention structures can be added to test generality and failure modes. Table~\ref{tab:related_survey} groups representative encrypted-traffic systems by operating model and evaluation axes that matter for consumer gateway comparability.

\begin{table*}[t]
\centering
\caption{Comparison of Encrypted-Traffic Security Systems for Stand-Alone Gateway Deployment}
\label{tab:related_survey}
\scriptsize
\setlength{\tabcolsep}{3.0pt}
\renewcommand{\arraystretch}{1.12}
\begin{tabular}{p{2.45cm}p{2.25cm}p{2.45cm}p{0.8cm}p{2.25cm}p{2.25cm}p{2.95cm}}
\toprule
\textbf{Group / work} &
\textbf{Inputs} &
\textbf{Training \& calibration} &
\textbf{Online} &
\textbf{Robustness / evasion} &
\textbf{Action / impact} &
\textbf{Fit to NOS-Gate assumptions} \\
\midrule

\multicolumn{7}{l}{\textit{G0. Gateway action loop (detection $\rightarrow$ mitigation $\rightarrow$ queue impact)}}\\
NOS-Gate (this paper) &
Per-flow metadata + contention signals &
Label-free burn-in; per-flow high-quantile thresholds; $K$-of-$M$ persistence &
Yes &
Explicit timing-control budgets; feasibility reported &
WFQ weight gating; queue-tail \& collateral via replay &
Direct match: stand-alone calibration and action-impact loop \\

\addlinespace[1mm]
\multicolumn{7}{l}{\textit{G1. Interaction-aware encrypted-traffic detection (detection-first)}}\\
HyperVision \cite{fu23hypervision} &
Encrypted traffic features + flow-interaction graph &
Dataset-trained; calibration follows paper/dataset protocol &
Yes &
Targets unknown/novel threats (not posed as timing-budget shaping) &
No mitigation or queue impact &
Closest on online inference, but different calibration and threat model \\

\addlinespace[1mm]
\multicolumn{7}{l}{\textit{G2. Robust learning under weak supervision / data quality shift}}\\
RAPIER \cite{qing24rapier} &
Encrypted traffic detection features &
Robust training under low-quality data; typically offline &
Often &
Robustness to data quality shift (rather than timing budgets) &
No action loop &
Useful robustness angle, but not an action-coupled gateway protocol \\

\addlinespace[1mm]
\multicolumn{7}{l}{\textit{G3. Evasion emphasis (action-free evaluation)}}\\
Wedjat \cite{gao25wedjat} &
Encrypted traffic signals + causal analysis components &
Dataset-driven learning and evaluation protocol &
Yes &
Explicit evasion emphasis; real-time framing &
No mitigation or queue impact &
Threat focus is closer, but lacks stand-alone calibration and impact loop \\

\addlinespace[1mm]
\multicolumn{7}{l}{\textit{G4. Self-supervised representation learning (pretrain then downstream scoring)}}\\
ET-SSL \cite{sattar25etssl} &
Encrypted traffic features &
Self-supervised pretraining; downstream calibration varies by use case &
Often &
Not formulated as timing-budget shaping; robustness depends on downstream task &
No action loop &
Promising feature learning, but comparability hinges on calibration and impact evaluation \\

\bottomrule
\end{tabular}
\end{table*}

\section{System model (consumer gateway)}
\label{sec:system-model}

We model a stand-alone consumer gateway that observes packet metadata behind NAT and produces streaming decisions at a fixed cadence. Time is divided into non-overlapping windows of length \(\Delta t=\SI{250}{ms}\) (Section~\ref{sec:tm-observation}). The detector operates on directed five-tuple flows and is calibrated without labels using an assumed-benign burn-in segment. \rev{ Flows may start and stop during a horizon. For a newly observed flow, we compute scores immediately, but we enable thresholded alarming only after the burn-in segment needed for quantile calibration has accrued; before that, we suppress alarms and do not allow WFQ gating.}

\paragraph*{Inputs and operating protocol.}
The feature stream is limited to fields visible at the gateway: packet headers, timestamps, and sizes, plus DNS and TLS handshake metadata when present (for example, SNI and certificate validity days). We do not use payload bytes, deep packet inspection features, or host logs. If DNS/TLS fields are absent, NOS-Gate falls back to timing and size features without changing the model. To avoid mixing feature dimensions within a run, we evaluate a fixed feature contract per run (for example, timing+contention only, or timing plus DNS/TLS when uniformly available). The run manifest records the feature contract used.

For each flow \(i\) and window \(t\), the detector outputs a score \(s_{i,t}\) and an alarm
\(a_{i,t}=\1\{s_{i,t}\ge \vartheta_i\}\).

\rev{Thresholds are calibrated label-free from the burn-in segment. Let the horizon be \(H\) windows and let the burn-in length be \(H_{\mathrm{burn}}=\lfloor 0.6H\rfloor\). In our worlds, \(H=57{,}600\) and \(H_{\mathrm{burn}}=34{,}560\) windows at \(\Delta t=\SI{250}{ms}\) (about \(2.4\) hours). We set
\begin{equation}
\vartheta_i \;=\; \mathrm{Quantile}_q\Bigl(\{s_{i,t}: t=0,\dots,H_{\mathrm{burn}}-1\}\Bigr),
\label{eq:quantile_threshold}
\end{equation}
with \(q\in\{0.99,0.999\}\) as the operating knob (we report both). Before \(\vartheta_i\) is available, we compute \(s_{i,t}\) causally but suppress alarms (\(a_{i,t}=0\)) and do not trigger gating; after burn-in, thresholds are frozen and performance is reported on the held-out segment, including achieved benign false-positive rate.}

To avoid flapping, we apply a \(K\)-of-\(M\) persistence rule. A flow becomes actionable only if at least \(K\) alarms occur in the most recent \(M\) windows (defaults \(K=3\), \(M=8\)):
\begin{equation}
z_{i,t} \;=\; \1\left\{\sum_{\tau=t-M+1}^{t} a_{i,\tau} \ge K\right\},
\end{equation}

\rev{where the actionable indicator $z_{i,t}$ is defined by the rolling $K$-of-$M$ rule above. WFQ gating is implemented as a reversible quarantine timer: when $z_{i,t}=1$ for a flow, we apply $\omega_i=\omega_-$ for a fixed quarantine duration $T_g$. If further actionable triggers occur while a flow is already gated, the quarantine interval is extended accordingly. If no further triggers occur, the flow returns automatically to $\omega_0$ when the quarantine timer expires.}

\begin{figure}[t]
\centering
\begin{tikzpicture}[
  font=\fontsize{7pt}{7pt}\selectfont,
  >=Latex,
  node distance=2mm,
  box/.style={draw, rounded corners, inner sep=4pt, line width=0.5pt},
  blk/.style={draw, rounded corners, align=center, minimum height=7mm, inner sep=2pt, fill=white},
  lab/.style={font=\scriptsize, align=center},
  line/.style={-Latex, thick},
  dline/.style={-Latex, thick, dashed}
]

\pgfdeclarelayer{bg}
\pgfsetlayers{bg,main}

\node[blk, minimum width=32mm] (wfq) {WFQ scheduler};
\node[blk, below=3mm of wfq, minimum width=32mm] (qmet)
{Queue state\\{\scriptsize backlog, service share}};
\node[box, fit=(wfq)(qmet), inner sep=3pt,
  label={[lab, yshift=1mm]above:\textbf{Scheduling domain}}] (schedbox) {};

\node[blk, below=8mm of schedbox, minimum width=32mm] (feat)
{Windowing + features\\{\scriptsize (metadata-only)}};
\node[blk, below=3mm of feat, minimum width=32mm] (nos)
{NOS-Gate state update\\{\scriptsize state $(v,u)$, $S=\sigma_k(v)$}};
\node[blk, below=3mm of nos, minimum width=32mm] (pers)
{Threshold $\vartheta_i$\\{\scriptsize burn-in quantile}\\{\scriptsize + persistence $K$-of-$M$}};
\node[blk, below=3mm of pers, minimum width=24mm] (alarm)
{Actionable $z_{i,t}$};
\node[box, fit=(feat)(nos)(pers)(alarm), inner sep=3pt,
  label={[lab, yshift=1mm]above:\textbf{Gateway runtime}}] (gwbox) {};

\node[blk, below=8mm of gwbox, minimum width=32mm] (worlds)
{Worlds generator\\{\scriptsize budgets $(R^{\min},\varepsilon,\delta_q)$}};
\node[blk, below=3mm of worlds, minimum width=28mm] (trace)
{Packet trace};
\node[blk, below=3mm of trace, minimum width=32mm] (replay)
{WFQ replay};

\node[blk, below=3mm of replay, minimum width=32mm] (metrics)
{Report\\{\scriptsize FPR, recall, TTD}\\{\scriptsize p99/p99.9, collateral}\\ (Labels for reporting only)};

\node[box, dashed, fit=(worlds)(trace)(replay)(metrics), inner sep=3pt,
  inner ysep=8pt,
  label={[lab, yshift=1mm]above:\textbf{Offline evaluation}}] (offbox) {};

\coordinate (midpoint) at ($(schedbox.south)!0.5!(offbox.north)$);

\node[blk,
      text width=24mm,
      text height=4mm,
      align=center,
      rotate=270,
      right=18mm of midpoint,
      xshift=2mm,
      yshift=12mm] (traffic)
      {Traffic (flows)};

\node[lab, above=-1mm of traffic, font=\tiny, rotate=270]
  {\textit{timestamps, sizes, headers}};

\begin{pgfonlayer}{bg}
  \node[draw=none, rounded corners, fit=(schedbox)(gwbox),
        inner sep=5pt, fill=black!25, fill opacity=0.18] {};
  \node[draw=none, rounded corners, fit=(offbox),
        inner sep=5pt, fill=gray!35, fill opacity=0.18] {};
\end{pgfonlayer}


\draw[line] (wfq) -- (qmet);

\draw[line] (traffic.west) to[out=90, in=0]
  node[lab, above, pos=0.6, font=\tiny, xshift=-4mm] {packets} (wfq.east);

\draw[line] (traffic.south) to[out=180, in=0]
  node[lab, below, pos=0.9, font=\tiny, xshift=4mm, yshift=4mm] {packets} (feat.east);

\draw[line] (wfq.south west) to[out=240, in=180]
  node[lab, right, pos=0.55, font=\tiny, text width=12mm, xshift=-3mm] {contention\\signals}
  (feat.west);

\draw[line] (alarm.west) to[out=180, in=180, looseness=0.6]
  node[lab, left, pos=0.8, font=\tiny, text width=14mm] {$\omega_0 \!\to\! \omega_-$ for $T_g$}
  (wfq.west);

\draw[dline] (worlds) -- (trace);
\draw[dline] (trace) -- (replay);

\draw[dline] (trace.east) to[out=0, in=-90]
  node[lab, right, pos=0.55, font=\tiny, text width=12mm] {Trace-driven\\traffic}
  (traffic.east);

\draw[dline] (alarm.south east) to[out=-30, in=30]
  node[lab, right, pos=0.85, font=\tiny] {gating log} (replay.north east);

\draw[dline] (replay.south) -- (metrics.north);

\end{tikzpicture}
\caption{Online, the gateway windows metadata-only traffic, updates NOS-Gate state \((v,u)\), applies burn-in quantile thresholding and a \(K\)-of-\(M\) persistence rule to produce an actionable flag \(z_{i,t}\), and temporarily reduces WFQ weight \(\omega_0\!\to\!\omega_-\) for \(T_g\). Offline, \emph{worlds} generates trace-driven traffic and labels, runs the same runtime to log gating decisions, and replays WFQ to report detection metrics and queue-tail and collateral outcomes; labels are used only for reporting. Solid arrows: deployed online loop. Dashed arrows: offline evaluation harness.}
\label{fig:system_overview}
\end{figure}
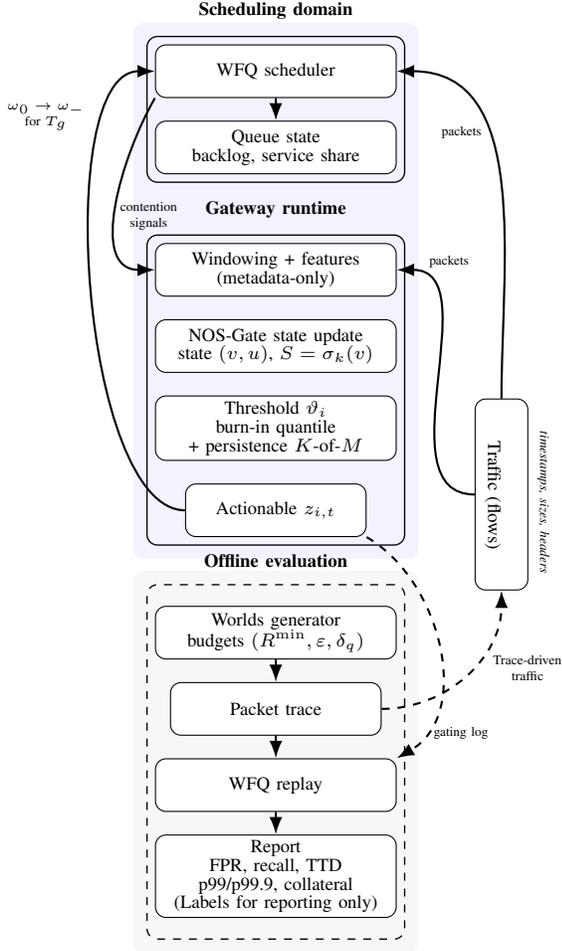

\paragraph*{Queueing model and WFQ action.}
We evaluate action impact through packet-level queue replay at the gateway. Flows that share the uplink enter a work-conserving weighted fair queueing (WFQ) scheduler. Let \(\omega_i(t)\) be the weight assigned to flow \(i\) at time \(t\). Under normal operation, \(\omega_i(t)=\omega_0\). Under gating, a flow that is actionable (\(z_{i,t}=1\)) is deprioritised by setting \(\omega_i(t)=\omega_-\) with \(\omega_-<\omega_0\) for a quarantine duration \(T_g\) (or until the flag clears, whichever is longer). We report both detection metrics and queue-tail metrics, including collateral delay experienced by benign flows. WFQ gating is reversible and avoids immediate blocking while still limiting queue-tail harm when alarms persist (Figure~\ref{fig:system_overview}).
\rev{
\paragraph*{Runtime measurability and clique approximation.}
In \emph{worlds}, contention cliques and weights \(W\) are specified by construction to make interference and \(\rho(W)\) auditable. In a consumer gateway, clique membership is not given as an explicit graph label, but the relevant observables are available at the scheduler boundary: per-queue backlog/occupancy (or delay estimates), per-class service counters, and WFQ configuration/state (weights and per-class rate share). Operationally, we approximate a “clique” as the set of flows that share the same constrained scheduling domain, such as a WAN egress queue and traffic class. Per-flow membership is obtained from the gateway’s own flow-to-class mapping (five-tuple classification or policy rules). This is sufficient for the contention features we use (\texttt{share}, \texttt{q}) and for implementing reversible weight gating without requiring an online oracle for \(W\).}

\subsection{Worlds benchmark: contract and outputs}
\label{sec:worlds-contract}

The primary evidence in this paper is produced by a reproducible synthetic benchmark called \emph{worlds}. A \emph{world} is a fully specified, executable instance that fixes benign device processes, malicious episode generators, contention structure, and an evaluation harness for queue impact. Each world publishes: (i) benign device classes and traffic generators, (ii) malicious episode generators, (iii) contention structure (cliques and weights \(W\), including a target band for \(\rho(W)\)), (iv) a packet trace with flow keys, (v) the derived windowed feature stream, (vi) labels used only for reporting, and (vii) WFQ replay configuration and queue summaries. The released artefacts include world ids, random seeds, configuration hashes, and manifests linking raw and derived outputs.

\begin{table}[t]
\centering
\color{black}
\caption{Scale of the \emph{worlds} benchmark used in this paper. Offered-load statistics are derived from benign generator parameters (mean IAT family parameters and mean packet sizes); realised traces vary stochastically around these means.}
\label{tab:worlds_scale}
\scriptsize
\setlength{\tabcolsep}{4pt}
\renewcommand{\arraystretch}{1.12}
\begingroup
\begin{tabular}{p{3.2cm}p{3.8cm}}
\toprule
Quantity & Value (this paper) \\
\midrule
Window length $\Delta t$ & $\SI{250}{ms}$ \\
Horizon $H$ & $57{,}600$ windows ($4$ hours) \\
Devices / directed flows & $32$ devices; $64$ directed flows ($2$ per device) \\
Malicious flows / episodes & $3$ malicious flows per world (active at any given time) \\
Clique structure & $4$ cliques, size $16$ each \\
Contention density & $0.25$ (world construction setting) \\
$\rho(W)$ band (achieved) & $\approx 0.90$ per clique (by construction; narrow band around target) \\
Per-flow rate (pkts/s) & min / median / max $\approx 0.16 / 4.68 / 14.15$ \\
Per-flow offered load (kb/s) & min / median / max $\approx 0.61 / 15.32 / 60.85$ \\
Total offered load (Mb/s) & $\approx 1.59$ (sum over 64 flows) \\
\bottomrule
\end{tabular}
\endgroup
\end{table}
\paragraph*{Budgets, feasibility, and timeline.}
Timing-controlled evasion is defined by auditable operational budgets \((R_i^{\min},\varepsilon,\delta_q)\) and a feasibility protocol. We define these budgets in the threat model (Section~\ref{sec:tm-budgets}, Table~\ref{tab:budgets}) and report feasibility rates for each budget setting. Within each world, time is split chronologically into burn-in, validation, and test. During burn-in we update the online normaliser, train label-free baselines, and collect burn-in scores to set per-flow thresholds \(\vartheta_i\). Validation is used only for design-time selection from a small fixed hyperparameter grid. Test uses frozen detectors and frozen \(\vartheta_i\) to report achieved false-positive rate, recall, detection delay, and queue-tail outcomes with and without gating.

\section{Threat model}
\label{sec:threat-model}

We consider a stand-alone consumer gateway that observes packet metadata behind NAT and emits streaming, flow-level decisions at a fixed window cadence. The attacker can reshape packet timing and, within bounds, packet sizes. Evasion is constrained by explicit operational budgets that make attacks auditable in our \emph{worlds} benchmark.

\subsection{Observation model and feature access}
\label{sec:tm-observation}

Packets are indexed by \(n\), with arrival times \(t_n \in \RR_{\ge 0}\) and lengths \(\ell_n \in [\ell_{\min},\ell_{\max}]\).
A directed flow \(i\) is the five-tuple
\((\mathrm{srcIP},\mathrm{dstIP},\mathrm{srcPort},\mathrm{dstPort},\mathrm{proto})\).
Time is partitioned into non-overlapping windows of length \(\Delta t\), indexed by
\(t \in \{0,1,\dots,H-1\}\).

Let \(N_{i,t}\) be the number of packets from flow \(i\) in window \(t\),
\begin{equation}
N_{i,t} \;=\; \left|\left\{n \in i : t_n \in [t\Delta t,(t+1)\Delta t)\right\}\right|.
\end{equation}
When \(N_{i,t}\ge 2\), define the inter-arrival-time (IAT) multiset within the window as
\begin{equation}
\mathcal{I}_{i,t} \;=\; \left\{t_{n+1}-t_n : n,n+1 \in i,\; t_n,t_{n+1} \in [t\Delta t,(t+1)\Delta t)\right\}.
\end{equation}
The detector receives a feature vector \(x_{i,t}\in\RR^d\) computed causally from packet timestamps and sizes, optional DNS/TLS handshake metadata when visible, and contention signals available at the gateway (Section~\ref{sec:preprocess}). No payload content and no host logs are used.

\subsection{Benign traffic and malicious episodes}
\label{sec:tm-benign-malicious}

Within a world, each flow \(i\) is generated by either a benign process or a malicious episode generator.
Let \(\mathbb{P}_i^{\mathrm{ben}}\) and \(\mathbb{P}_i^{\mathrm{mal}}\) denote the per-window distributions of \(x_{i,t}\) in benign and malicious modes.
Flows that share the same uplink scheduling domain are grouped into contention cliques. We represent clique-level interference using a nonnegative weight matrix \(W=[w_{ij}]\), and we report its spectral radius \(\rho(W)\) as a compact summary of contention intensity.

\subsection{Adversary capabilities and knowledge}
\label{sec:tm-adversary}

We assume a remote adversary that can compromise IoT devices and generate malicious traffic while remaining behind NAT. The adversary can: (i) use encrypted channels, (ii) choose protocols and endpoints consistent with common device behaviour, and (iii) reshape packet timing and, within bounds, packet sizes. The adversary knows the feature contract and the window length \(\Delta t\).

The defender calibrates thresholds in a label-free, stand-alone way during an assumed-benign burn-in segment (Section~\ref{sec:system-model}). This reflects a standard operational assumption: burn-in is not dominated by malicious behaviour.

\subsection{Operational budgets for timing-controlled evasion}
\label{sec:tm-budgets}

Evasion is constrained by three budgets that capture objective, timing stealth, and contention stealth. For a flow \(i\), the budgets are \((R_i^{\min},\varepsilon,\delta_q)\) and are enforced by the worlds generator (Table~\ref{tab:budgets}).

For timing stealth, each malicious flow \(i\) is paired with a benign IAT reference distribution \(P^{\mathrm{ben,iat}}_i\) constructed within the same world, using the benign process that matches \(i\)'s device class and endpoint template. This makes the timing constraint auditable per world. \rev{For readability, we use $P^{\mathrm{ben,iat}}_i$ to denote the (fixed) benign IAT reference distribution for flow $i$, and $\widehat P^{\mathrm{iat}}_{i,t}$ for the empirical IAT distribution observed within window $t$; this notation is used consistently throughout the threat model and Algorithm~\ref{alg:evasivec2}.}

IAT-based distortion is undefined when \(N_{i,t}<2\). We therefore apply the timing distortion constraint only on windows that contain at least two packets:
\begin{equation}
\mathcal{T}_i \;=\; \{t \in \{0,\dots,H-1\} : N_{i,t}\ge 2\}.
\end{equation}
This does not remove constraints in sparse windows. The detector still observes packet counts, rates, and size-based features in every window. The restriction applies only to the IAT distortion term because it is undefined otherwise.

For contention stealth, let \(C(i)\) be the contention clique containing flow \(i\). Under baseline WFQ replay \emph{without defensive gating}, let \(d_n\) be the queueing delay (enqueue-to-dequeue time) experienced by packet \(n\) in clique \(C\). Define the clique mean delay over the horizon as
\begin{equation}
\overline{D}_C \;=\; \frac{1}{|\mathcal{P}_C|}\sum_{n\in\mathcal{P}_C} d_n,
\end{equation}
where \(\mathcal{P}_C\) is the set of packets belonging to flows in clique \(C\) during \([0,H\Delta t)\).
We write \(\overline{D}^{\mathrm{ben}}_C\) and \(\overline{D}^{\mathrm{atk}}_C\) for the mean delays under benign-only and attack-injected traces, respectively.

\begin{table}[t]
\centering
\caption{Operational budgets defining timing-controlled evasion in \emph{worlds}.}
\label{tab:budgets}
\scriptsize
\setlength{\tabcolsep}{4pt}
\begin{tabular}{p{1.7cm}p{0.7cm}p{0.6cm}p{4.2cm}}
\toprule
Budget & Symbol & Units & Definition \\
\midrule
Throughput floor
& $R_i^{\min}$
& bytes
& $\sum_{n\in i,\; t_n \in [0,H\Delta t)} \ell_n \ge R_i^{\min}$ \\

Timing distortion
& $\varepsilon$
& s
& $\frac{1}{|\mathcal{T}_i|}\sum_{t\in\mathcal{T}_i}
   W_1(\widehat P^{\mathrm{iat}}_{i,t}, P^{\mathrm{ben,iat}}_i)\le \varepsilon$ \\

Contention stealth
& $\delta_q$
& s
& $\overline{D}^{\mathrm{atk}}_{C(i)}-\overline{D}^{\mathrm{ben}}_{C(i)} \le \delta_q$ \\
\bottomrule
\end{tabular}
\end{table}

Some triples \((R_i^{\min},\varepsilon,\delta_q)\) are infeasible for a given world and clique load, so feasibility is treated as a first-class outcome. When a requested triple is infeasible, the generator either declares infeasibility under the requested budgets or attempts to restore feasibility by reducing offered load while preserving the throughput floor when possible. If the throughput floor cannot be met jointly with \((\varepsilon,\delta_q)\), the attempt is declared infeasible. We report feasibility rates for each budget setting.

\subsection{Defender evaluation under stand-alone thresholding}
\label{sec:tm-defender}

For flow \(i\) and window \(t\), the detector outputs a score \(s_{i,t}\) and a binary alarm
\begin{equation}
a_{i,t}=\1\{s_{i,t}\ge \vartheta_i\},
\end{equation}
where \(\vartheta_i\) is set by high-quantile burn-in thresholding (Eq.~\eqref{eq:quantile_threshold}).
Let \(\mathcal{A}(R_i^{\min},\varepsilon,\delta_q)\) denote the admissible adversarial policies that satisfy Table~\ref{tab:budgets}.
For a fixed quantile \(q\), we evaluate robustness through the achieved detection probability under admissible attacks,
\begin{equation}
\min_{P\in \mathcal{A}(R_i^{\min},\varepsilon,\delta_q)}
\Pr_{x\sim P}\!\left[s(x)\ge \vartheta_i \mid \mathrm{mal}\right],
\end{equation}
and we report the achieved benign false-positive rate induced by the same \(\vartheta_i\) on the held-out test segment.

\subsection{Evasive policy class: projection-based C2}
\label{sec:tm-evasion}

To avoid cherry-picked evasions, we use a fixed, auditable policy class that directly enforces the three budgets.
The generator starts from a malicious proposal schedule, then (i) projects timing within each window to satisfy the Wasserstein constraint, (ii) repairs sizes to maintain the throughput floor, and (iii) enforces contention stealth through iterative load reduction with baseline WFQ replay. \rev{To satisfy the three budgets, we apply the projection-based procedure summarized in Algorithm~\ref{alg:evasivec2}. The term \emph{budget-projection} refers to taking an initial malicious packet schedule and projecting it into the feasible set defined by $(R_i^{\min},\varepsilon,\delta_q)$ via an order-preserving time warp (quantile mapping of within-window IATs) plus local size repair to maintain the throughput floor.} \rev{
Here “projection” is used in the optimisation sense: starting from a proposed malicious schedule, we apply an order-preserving time-warp within each window (implemented as quantile mapping of sorted IATs with minimal interpolation) to satisfy the Wasserstein radius constraint, then repair sizes to maintain the throughput floor and iterate load reduction to satisfy contention stealth.}

\begin{algorithm}[h!]
\caption{Budget-projection evasive C2 generator (per flow)}
\label{alg:evasivec2}
\begin{algorithmic}[1]
\Require Benign IAT reference $P^{\mathrm{ben,iat}}_i$, size bounds $[\ell_{\min},\ell_{\max}]$, window size $\Delta t$, horizon $H$,
bytes floor $R_i^{\min}$, distortion radius $\varepsilon$, delay tolerance $\delta_q$, max iters $I_{\max}$
\Ensure Packet times $\{t_n\}$ and sizes $\{\ell_n\}$ for flow $i$ over $[0,H\Delta t)$, or infeasibility flag
\State \textbf{Initialise} a proposal schedule with packet times and sizes satisfying $\sum_{n} \ell_n \ge R_i^{\min}$
\For{$t=0$ to $H-1$}
  \State Extract packets in window $t$ and compute $N_{i,t}$
  \If{$N_{i,t}\ge 2$}
    \State Compute $\widehat P^{\mathrm{iat}}_{i,t}$ and $d_t \gets W_1(\widehat P^{\mathrm{iat}}_{i,t}, P^{\mathrm{ben,iat}}_i)$
    \If{$d_t > \varepsilon$}
      \State \textbf{Project timing in window:} apply an order-preserving time-warp on $[t\Delta t,(t+1)\Delta t)$
      \Statex \hspace{2.2em} so the resulting IAT empirical distribution satisfies $W_1\le \varepsilon$
      \Statex \hspace{2.2em} (implementation: quantile mapping on sorted IATs with minimal interpolation)
    \EndIf
  \EndIf
  \State \textbf{Repair throughput locally:} adjust packet sizes within $[\ell_{\min},\ell_{\max}]$ to keep cumulative bytes on track for $R_i^{\min}$
\EndFor
\State Run baseline WFQ replay (no gating) for clique $C(i)$ to estimate $\overline{D}^{\mathrm{atk}}_{C(i)}$
\For{$\ell=1$ to $I_{\max}$}
  \If{$\overline{D}^{\mathrm{atk}}_{C(i)}-\overline{D}^{\mathrm{ben}}_{C(i)} \le \delta_q$}
    \State \Return packet times and sizes
  \EndIf
  \State \textbf{Reduce offered load:} decrease packet emission while attempting to preserve $\sum_n \ell_n \ge R_i^{\min}$
  \State Re-apply the per-window projection and size repair steps
  \State Re-estimate $\overline{D}^{\mathrm{atk}}_{C(i)}$ by baseline WFQ replay
\EndFor
\State \Return \textbf{infeasible} under $(R_i^{\min},\varepsilon,\delta_q)$
\end{algorithmic}
\end{algorithm}

\subsection{Optional protocol side-channel constraints}
\label{sec:tm-protocol-constraints}

\rev{When DNS or TLS handshake metadata is available, we include the DNS/TLS features listed in Table~\ref{tab:feature_contract} as part of the fixed contract for that run. When such metadata is not present for a particular flow-window, the corresponding DNS/TLS feature values default to zero under the fixed contract. We also report results under a timing+contention-only contract (i.e., without DNS/TLS features) when those fields are not available in a uniform way. The run manifest records the feature contract and discrete settings (including $B$ and the z-score clip magnitude).}

\section{NOS-Gate detector: mapping IDS to NOS dynamics}
\label{sec:nos-detector}

NOS-Gate instantiates the NOS two-state unit as a streaming intrusion detector at a consumer gateway. The
two-state dynamics, bounded excitability, recovery coupling, and damping terms are inherited from NOS. The
change is the meaning of the exogenous drive: in NOS networking it represents queue and interference forcing,
whereas in NOS-Gate it is a metadata-only evidence signal computed per flow per window. This retains the NOS
structure while changing what the ``input'' represents.

Table~\ref{tab:nos_to_nosgate} provides an auditable correspondence between the NOS symbols and their IDS
interpretation, and makes clear what is reused unchanged and what is reinterpreted for detection and action.

\subsection{Design rationale: why a two-state NOS unit helps under timing control}
\label{sec:nos-rationale}

NOS-Gate targets the gateway loop in Figure~\ref{fig:system_overview}: flows must be scored online at a fixed
cadence, thresholds must be set without labels, and decisions should drive a reversible action with measurable
queue impact. Under the threat model, the attacker can time-warp within windows and reshape burst structure
subject to a throughput floor and a timing-distortion budget. This often reduces sharp per-window signatures and
pushes attacks toward smaller deviations that persist over longer periods.

Stand-alone thresholding makes this harder. Each flow threshold \(\vartheta_i\) is set by a high burn-in quantile
(Eq.~\eqref{eq:quantile_threshold}) and then frozen. A detector that relies on rare, high-amplitude score spikes
is easier to suppress because the attacker only needs to keep scores below a tail threshold. A detector that can
accumulate modest deviations over time is more likely to cross a high quantile without requiring any single
window to look extreme.

With coupling disabled by default (\(g=0\)), NOS-Gate acts as a per-flow streaming mechanism driven by the scalar
evidence \(E_{i,t}\) in Eq.~\eqref{eq:nosgate_evidence}. Around the benign operating band, the \(v\)-update behaves
like a leaky evidence accumulator, while the recovery state \(u\) provides slower negative feedback. The leak
terms prevent unbounded accumulation, the bounded nonlinearity reduces sensitivity to rare transients, and the
explicit \(-u\) term suppresses repeated triggering after sustained elevation. The event surrogate
\(S_{i,t}=\sigma_k(v_{i,t})\) converts this filtered state into a bounded readout that is straightforward to
calibrate using a high-quantile rule.

\subsection{From gateway flows to NOS nodes}
\label{sec:nosgate_nodes}

We model each active directed flow \(i\) as a node in a contention structure with nonnegative weights
\(W=[w_{ij}]\). Two flows are neighbours if they compete for the same gateway scheduling domain (uplink WFQ
domain or a contention clique). The NOS graph-local input is retained:
\begin{equation}
I_{i,t} \;=\; g\sum_{j\in\mathcal{N}(i)} w_{ij}\,S_{j,t-L_{ij}},
\label{eq:nosgate_input}
\end{equation}
where \(g\) is a global coupling gain, \(L_{ij}\in\mathbb{Z}_{\ge 0}\) is an integer delay in windows, and
\(S_{j,t}\) is an event surrogate emitted by flow \(j\) (defined below). In the main results we set \(g=0\) unless
stated otherwise; coupling is evaluated in ablations.
\rev{We update all flows synchronously once per window. Since the neighbour surrogate $S_{j,t}$ is emitted at window $t$ and is only available when advancing to $t{+}1$, the smallest realised coupling lag is one full window. We therefore parameterise an \emph{additional} delay $\tau_{ij}\in\mathbb{Z}_{\ge 0}$ beyond this inevitable pipeline delay and define $L_{ij}=1+\tau_{ij}$, so $\tau_{ij}=0$ corresponds to the minimum one-window lag.}
\subsection{State interpretation for IDS}
\label{sec:nosgate_state}

NOS-Gate reuses the two NOS states with an IDS interpretation that preserves their mathematical roles:
\[
\begin{aligned}
v &\mapsto \text{bounded evidence accumulation (suspicion)},\\
u &\mapsto \text{recovery state (suppression resource)}.
\end{aligned}
\]
Operationally, \(v\) behaves like a virtual evidence queue that rises when metadata features deviate from benign
patterns and drains under damping. The recovery state \(u\) increases under sustained elevation and pulls \(v\)
down via the explicit \(-u\) term, which reduces rapid re-triggering after a burst of alarms. When we include \(u\)
in a score readout, we treat it as a persistence indicator correlated with sustained deviation, not as a direct
alarm trigger by itself.

\subsection{Evidence drive as a magnitude input}
\label{sec:nosgate_evidence}

For each flow $i$ and window $t$, we compute a normalised feature vector $\hat{x}_{i,t}\in\RR^d$
(Section~\ref{sec:preprocess}) and map it to a scalar evidence drive
\begin{equation}
E_{i,t} \;=\; \zeta \,\|\hat{x}_{i,t}\|_p,
\label{eq:nosgate_evidence}
\end{equation}
where $p\ge 1$ and $\zeta>0$ are hyperparameters. We treat $E_{i,t}$ as constant over the window
$[t\Delta t,(t+1)\Delta t)$. Because $\hat{x}_{i,t}$ is produced by online z-scoring, the norm aggregates
multi-feature deviations without labels and avoids learning per-feature weights.

\subsection{NOS dynamics for IDS (windowed discretisation)}
\label{sec:nosgate_dynamics}

NOS-Gate runs once per window, so we discretise the NOS two-state dynamics with an explicit Euler step of size
$\Delta t$. Let $v_{i,t}\approx v_i(t\Delta t)$ and $u_{i,t}\approx u_i(t\Delta t)$.

\textbf{Event surrogate (readout used for coupling and scoring).}
\begin{equation}
S_{i,t} \;=\; \sigma_k(v_{i,t}-\theta) \;=\; \frac{1}{1+\exp\!\bigl(-k\,(v_{i,t}-\theta)\bigr)}.
\label{eq:nosgate_event}
\end{equation}

\textbf{Neighbour drive (optional).}
When coupling is enabled, the NOS graph-local term is Eq.~(\ref{eq:nosgate_input}).
We update all flows synchronously once per window. Since $S_{j,t}$ is produced at window $t$ and is only available
when advancing to $t{+}1$, the minimum realised coupling lag is one window. We therefore parameterise an additional
delay $\tau_{ij}\in\mathbb{Z}_{\ge 0}$ and implement $L_{ij}=1+\tau_{ij}$, so $\tau_{ij}=0$ corresponds to the minimum
one-window lag.

\textbf{Bounded excitability.}
We use the NOS bounded nonlinearity
\begin{equation}
f_{\mathrm{sat}}(v) \;=\; \frac{\alpha v^2}{1+\kappa v^2},
\label{eq:nosgate_fsat}
\end{equation}
with $\alpha>0$ and $\kappa>0$. For numerical safety, the implementation may clip $v$ to $[0,v_{\max}]$


\textbf{State update (same NOS core, with IDS-specific drive).}
Define the total exogenous drive as the sum of the IDS evidence and (optionally) the neighbour drive:
$A_{i,t}=E_{i,t}+I_{i,t}$ (set $I_{i,t}=0$ when coupling is disabled). The per-window updates are

\begin{align}
\begin{split}
            v_{i,t+1} = v_{i,t} + \Delta t\Bigl( f_{\mathrm{sat}}(v_{i,t}) + \beta v_{i,t} + \gamma - u_{i,t} + A_{i,t} \\
            - \lambda v_{i,t}- \chi\bigl(v_{i,t}-v_{\mathrm{rest}}\bigr)\Bigr)c\;+\; \xi_{i,t} \;-\; r\,S_{i,t},
\label{eq:nosgate_v}
\end{split}
\end{align}

\begin{align}
u_{i,t+1}
&= u_{i,t} + \Delta t\Bigl(
ab\,v_{i,t} - (a+\mu)\,u_{i,t}
\Bigr).
\label{eq:nosgate_u}
\end{align}
Here $\xi_{i,t}$ is an optional zero-mean noise term (set to $0$ unless enabled) and $v_{\mathrm{rest}}$ is the benign
baseline (we use $v_{\mathrm{rest}}=0$ unless stated). The optional term $-rS_{i,t}$ is a continuous reset pathway
(set $r=0$ when not used). This is the only structural difference from NOS: the ``input'' is now $A_{i,t}$, whose
IDS component is $E_{i,t}$ from Eq.~\eqref{eq:nosgate_evidence}.

\begin{table}[t]
\centering
\caption{Auditable mapping from NOS (main model) to NOS-Gate (IDS instantiation).}
\label{tab:nos_to_nosgate}
\begin{tabular}{p{1.9cm}p{4.5cm}p{5.5cm}}
\toprule
Symbol & NOS meaning & NOS-Gate (IDS) meaning \\
\midrule
\(v_i\) &
Normalised congestion proxy &
Bounded evidence accumulation, optionally clipped to \([0,v_{\max}]\) \\
\(u_i\) &
Recovery / slowdown resource &
Suppression state damping repeated alarms \\
\(f_{\mathrm{sat}}(v)\) &
Bounded excitability (finite buffers) &
Bounded evidence growth under deviation \\
\(\beta v_i+\gamma\) &
Residual slope and offset &
Bias and linear drift correction \\
\(-u_i\) &
Post-burst recovery drag &
Refractory suppression after alarms \\
\(-\lambda v_i\) &
Service-like drainage &
Evidence decay toward baseline \\
\(-\chi(v_i-v_{\mathrm{rest}})\) &
Small-signal damping &
Stabilisation around benign baseline \\
\(I_i(t)\) &
Graph-local coupling &
Contention-driven neighbour propagation (ablation) \\
\(S_i(t)\) &
Presynaptic event train &
Event surrogate \(S_{i,t}=\sigma_k(v_{i,t}-\theta)\) \\
\((a,b,\mu)\) &
Recovery sensitivity and decay &
Suppression build-up and persistence \\
Exogenous drive &
Telemetry forcing &
Total drive $A_{i,t}=E_{i,t}+I_{i,t}$ with $E_{i,t}=\zeta\|\hat{x}_{i,t}\|_p$ \\
Reset pathway &
Differentiable reset option &
Optional continuous reset \(-rS_{i,t}\) \\
Operational action &
Operator control &
WFQ gating: alarms reduce weight \(\omega_i\) \\
\bottomrule
\end{tabular}
\end{table}

\subsection{Score, alarm, and WFQ gating action}
\label{sec:nosgate_score_action}

NOS-Gate converts state into an anomaly score
\begin{equation}
s_{i,t} \;=\; \eta_1\,S_{i,t} \;+\; \eta_2\,u_{i,t}.
\label{eq:nosgate_score}
\end{equation}
An alarm is $a_{i,t}=\1\{s_{i,t}\ge \vartheta_i\}$, where $\vartheta_i$ is set per flow by burn-in quantile thresholding
(Eq.~\eqref{eq:quantile_threshold}). We stabilise alerts with the $K$-of-$M$ persistence rule
(Section~\ref{sec:system-model}). When actionable, WFQ gating assigns $\omega_i=\omega_-$ for $T_g$; otherwise
$\omega_i=\omega_0$.

In the main results we use $\eta_2=0$, so the operational score is driven by $S_{i,t}$. We use $\eta_2>0$ only in
ablations, where $u_{i,t}$ acts as a slow-burn persistence cue. The run manifest records $(\eta_1,\eta_2)$.

\subsection{Fixed points and stability}
\label{sec:nosgate_stability}

For constant drives (that is, $E_{i,t}\equiv E_i^\star$ and $I_{i,t}\equiv I_i^\star$), with $\xi_{i,t}=0$ and $r=0$,
a steady state $(v_i^\star,u_i^\star)$ of the underlying NOS core satisfies
\begin{equation}
0 \;=\; f_{\mathrm{sat}}(v_i^\star) + \beta v_i^\star + \gamma - u_i^\star
+ E_i^\star + I_i^\star
- \lambda v_i^\star - \chi\bigl(v_i^\star - v_{\mathrm{rest}}\bigr),
\label{eq:nosgate_fixedpoint}
\end{equation}
together with $ab\,v_i^\star=(a+\mu)\,u_i^\star$ from Eq.~\eqref{eq:nosgate_u}. (If the optional reset $r>0$ is enabled,
it adds an additional negative feedback term proportional to $S_i^\star=\sigma_k(v_i^\star-\theta)$ in the discrete-time
update in Eq.~\eqref{eq:nosgate_v}.)

For conservative calibration checks, we isolate the coupling gain. Since $\max_x \sigma_k'(x)=k/4$, the linearised
coupling path in the $v$-update is bounded by $\Delta t\,g\,\rho(W)\,k/4$. In experiments we choose $(g,k)$, $\Delta t$,
and the world-specific $\rho(W)$ band so that this bound remains below the local damping and recovery margin induced by
$(\lambda,\chi)$ and $(a,b,\mu)$, and the local slope of $f_{\mathrm{sat}}$ around the operating band. The IDS evidence
$E_{i,t}$ enters only as exogenous forcing.

\rev{We emphasise that the bound $\Delta t\,g\,\rho(W)\,k/4$ is used as a conservative heuristic sufficient check for selecting safe coupling settings in our discrete-time implementation; it is not a formal stability guarantee for the full nonlinear, driven system under arbitrary time-varying evidence inputs.}

\section{Methods}
\label{sec:preprocess}

NOS-Gate operates on per-flow packet metadata at a fixed cadence. We treat each directed five-tuple flow as a stream and summarise its packets in non-overlapping windows of length $\Delta t=\SI{250}{ms}$. Each window yields one feature vector $x_{i,t}$ for flow $i$ at window index $t$. In synthetic \emph{worlds}, flow keys and (when needed) device identifiers are known by construction. In the optional corpus mode, we extract the same directed five-tuple flows from PCAP or flow logs. Device identity is not required for scoring, but when a stable identifier exists (for example, a MAC address or a DHCP lease), we maintain separate normalisation statistics per device.

\subsection{Causal windowing and online normalisation}
All feature computation is causal. Features for window $t$ use only packets with timestamps in $[t\Delta t,(t{+}1)\Delta t)$, and any running statistics depend only on earlier windows. After computing $x_{i,t}$, we normalise online to reduce drift and make thresholds comparable across flows.

Let $d(i)$ be the device identifier for flow $i$ when available (otherwise a single global bucket). For each feature index $k$ and device bucket $d$, we maintain exponential running mean and variance,
\begin{align}
m_{t+1}^{(d,k)} &= (1-\lambda_m)m_t^{(d,k)} + \lambda_m x_t^{(d,k)}, \\
q_{t+1}^{(d,k)} &= (1-\lambda_v)q_t^{(d,k)} + \lambda_v\left(x_t^{(d,k)}-m_t^{(d,k)}\right)^2,
\end{align}
and emit a z-score
\begin{equation}
\hat{x}_t^{(d,k)} = \frac{x_t^{(d,k)}-m_t^{(d,k)}}{\sqrt{q_t^{(d,k)}+\epsilon}}.
\label{eq:zscore}
\end{equation}
Operationally, we compute $\hat{x}_{i,t}$ using $(m_t,q_t)$, score window $t$, and then update $(m_{t+1},q_{t+1})$ using $x_{i,t}$. We use separate rates $\lambda_m$ and $\lambda_v$ so that the variance adapts more cautiously than the mean under bursty traffic. If a window has no packets, rate features are set to zero and IAT summaries are marked as missing (the evidence drive in Eq.~\eqref{eq:nosgate_evidence} remains defined through other features). We clip z-scores to a fixed magnitude (default $8$) to prevent a single spike from dominating.
Thresholds are fixed after burn-in. To limit post burn-in drift, online mean and variance updates use a small adaptation rate and are clipped to prevent abrupt shifts. We also evaluated freezing normalisation after burn-in and observed similar qualitative trends, at the cost of slightly higher false positives; we leave systematic adversarial drift handling to future work.

\subsection{Feature contract}
The feature stream is deliberately small and restricted to fields visible at the gateway: headers, timestamps, and sizes, plus DNS and TLS handshake metadata when present. From timing and burstiness we include packet rate, simple inter-arrival summaries (mean and coefficient of variation when at least two packets exist), and a low-frequency pacing indicator computed from counts in $B$ fixed micro-bins within the window. From contention we include (i) clique-level rate share and (ii) an interference index derived from the contention weights $W$ used by the \emph{worlds} generator.

\rev{For the main results, we use a fixed $d=14$ feature contract per flow-window; Table~\ref{tab:feature_contract} lists the features, and Table~\ref{tab:nosgate_defaults} lists the defaults of NOS used in the main runs. The duty-cycle feature uses $B=10$ micro-bins within each window.}

\begin{table}[t]
\centering
\scriptsize
\setlength{\tabcolsep}{6pt}
\renewcommand{\arraystretch}{1.05}
\color{black}
\caption{Feature contract used in the main runs ($d=14$, $B=10$).}
\label{tab:feature_contract}
\begin{tabular}{p{2.6cm}p{6cm}}
\toprule
Group & Features \\
\midrule
Timing/volume & \texttt{rate}, \texttt{byte\_rate}, \texttt{mean\_iat}, \texttt{cv\_iat} \\
Size/burstiness & \texttt{mean\_size}, \texttt{small\_frac}, \texttt{duty\_cycle} ($B=10$) \\
Contention & \texttt{share}, \texttt{q} \\
DNS & \texttt{dns\_churn}, \texttt{dns\_ttl\_var}, \texttt{dns\_q} \\
TLS & \texttt{tls\_sni\_entropy}, \texttt{tls\_cert\_valid\_days} \\
\bottomrule
\end{tabular}
\end{table}

\begin{table}[t]
\centering
\color{black}
\caption{NOS-Gate parameters and defaults used in the main runs (see released config snapshot).}
\label{tab:nosgate_defaults}
\scriptsize
\setlength{\tabcolsep}{4pt}
\renewcommand{\arraystretch}{1.12}
\begin{tabular}{lcl}
\toprule
Parameter & Default & Meaning \\
\midrule
$\Delta t$ & 0.25  & window length \\
$k$ & 20 & sigmoid slope in $S_{i,t}=\sigma_k(v_{i,t}-\theta)$ \\
$\theta$ & 1.0 & sigmoid threshold \\
$\alpha$ & 1.0 & saturation numerator in $f_{\mathrm{sat}}(v)=\frac{\alpha v^2}{1+\kappa v^2}$ \\
$\kappa$ & 0.75 & saturation denominator scale \\
$\beta$ & 0.10 & linear term in $v$ update \\
$\gamma$ & 0.25 & offset in $v$ update \\
$\lambda$ & 0.35 & decay in $v$ update \\
$\chi$ & 0.10 & stabilisation around $v_{\mathrm{rest}}$ \\
$a,b,\mu$ & 1, 1, 0.01 & recovery build-up and decay (defaults when not overridden) \\
$p$ & 2 & evidence norm $\|\hat{x}_{i,t}\|_p$ (L2) \\
$\zeta$ & 1.0 & evidence scale (default feature scaling) \\
$v_{\max}$ & 3.0 & optional clipping bound when enabled \\
\bottomrule
\end{tabular}
\end{table}

\rev{We densify to a full $(\text{flow},\text{window})$ grid. If a flow has no packets in window $t$, rate and byte-rate are set to zero; IAT summaries are set to zero when fewer than two packets are present in the window (since IAT is undefined). In runs that include DNS/TLS features (Table~\ref{tab:feature_contract}), when DNS/TLS metadata is not present for a particular flow-window, the corresponding DNS/TLS feature values default to zero under the fixed contract. We also report results under a timing+contention-only contract (i.e., without DNS/TLS features) when those fields are not available in a uniform way. The run manifest records the feature contract and discrete settings (including $B$ and the z-score clip magnitude).}

\subsection{End-to-end loop and replay-based impact evaluation}
The online pipeline is identical in synthetic and corpus modes. The \emph{worlds} generator emits packet traces with flow keys, timestamps, and sizes, together with contention structure and evasion budgets (Table~\ref{tab:budgets}). We window packets to produce $x_{i,t}$, normalise online to obtain $\hat{x}_{i,t}$, and run NOS-Gate as a per-flow streaming update (Section~\ref{sec:nos-detector}).

All detectors, including baselines, follow the same stand-alone operating protocol. An assumed-benign burn-in segment is used for training (where applicable) and for collecting burn-in scores. Each flow threshold is then set as a high quantile of burn-in scores (Eq.~\eqref{eq:quantile_threshold}) and frozen for test. Labels are used only for reporting on held-out time.

Queue impact is evaluated by packet-level WFQ replay in two conditions: (i) baseline replay without gating and (ii) replay with gating driven by each method’s actionable flags $z_{i,t}$. Replay logs are produced in microseconds; we report queue metrics in milliseconds for readability. Efficiency is reported as microseconds per $(\text{flow},\text{window})$ record, since gateway cost scales primarily with active flows and the window cadence.

\rev{
We compare NOS-Gate against three lightweight label-free baselines trained on the same burn-in segment and calibrated with the same high-quantile thresholding protocol. KitNET is implemented as an ensemble of single-hidden-layer autoencoders over feature groups of size $\lceil\sqrt{d}\rceil$ (default grouping), with per-group hidden width set to $0.5$ of the group size and an aggregator autoencoder with hidden width $8$ operating on the vector of per-group reconstruction errors. Autoencoder is a compact one-hidden-layer MLP autoencoder $\mathrm{Linear}(d\!\rightarrow\!32)\!+\!\mathrm{ReLU}\!+\!\mathrm{Linear}(32\!\rightarrow\!d)$ trained by reconstruction loss. TinyGRU is a one-layer GRU forecaster with hidden size $24$ and lookback $8$ windows, followed by a linear head that predicts the next normalised feature vector; its anomaly score is the one-step prediction error. All learnable baselines minimise mean squared error (reconstruction for KitNET/autoencoder; one-step prediction for TinyGRU) using Adam (learning rates as in the released configuration), with a fixed small number of epochs. We do not use early stopping in the reported runs; dropout is disabled and weight decay is set to zero unless stated otherwise.}

\section{Results}
\label{sec:results}

We report results on reproducible worlds containing malicious episodes under the timing-controlled threat model. 
All methods operate in a stand-alone, label-free mode: per-flow thresholds are set from the burn-in segment using the same high-quantile rule and then frozen for test (Eq.~\eqref{eq:quantile_threshold}). We report two alert budgets, $\text{FPR}=1\%$ and $\text{FPR}=0.1\%$, and we always show the \emph{achieved} benign false-alarm rate on the held-out test segment under that protocol. \rev{
We report the \emph{achieved} benign false-positive rate induced by the frozen per-flow thresholds on the held-out test segment. Concretely, for each method we compute the fraction of benign flow-windows in the test segment that produce an alarm under $a_{i,t}=\1\{s_{i,t}\ge \vartheta_i\}$, where $\vartheta_i$ is set from the burn-in segment (Eq.~\eqref{eq:quantile_threshold}) and then frozen. Because this estimate aggregates over a very large number of benign flow-windows per world in our evaluation, small world-to-world differences can fall below the precision implied by typical rounded tables. We therefore state operating points by their target alert budgets and always report the achieved rate under the same protocol. In our harness, achieved FPR is estimated on a benign-only episode over the full horizon, which yields $64\times 57{,}600 \approx 3.69\times 10^6$ flow-windows per world at $\Delta t=\SI{250}{ms}$. At target $\mathrm{FPR}=0.1\%$, the sampling variability of an empirical false-positive fraction at this $n$ is on the order of $10^{-5}$, so values rounded to three decimals can appear identical across worlds.
For this reason we state operating points in terms of the target budget and clarify the estimation volume, and we report achieved values at higher precision in the artefact summaries.}

\subsection{Incident detection under timing-controlled evasion}
\begin{figure*}[h!]
\centering
\begin{minipage}{0.65\textwidth}
\centering
\includegraphics[width=\linewidth]{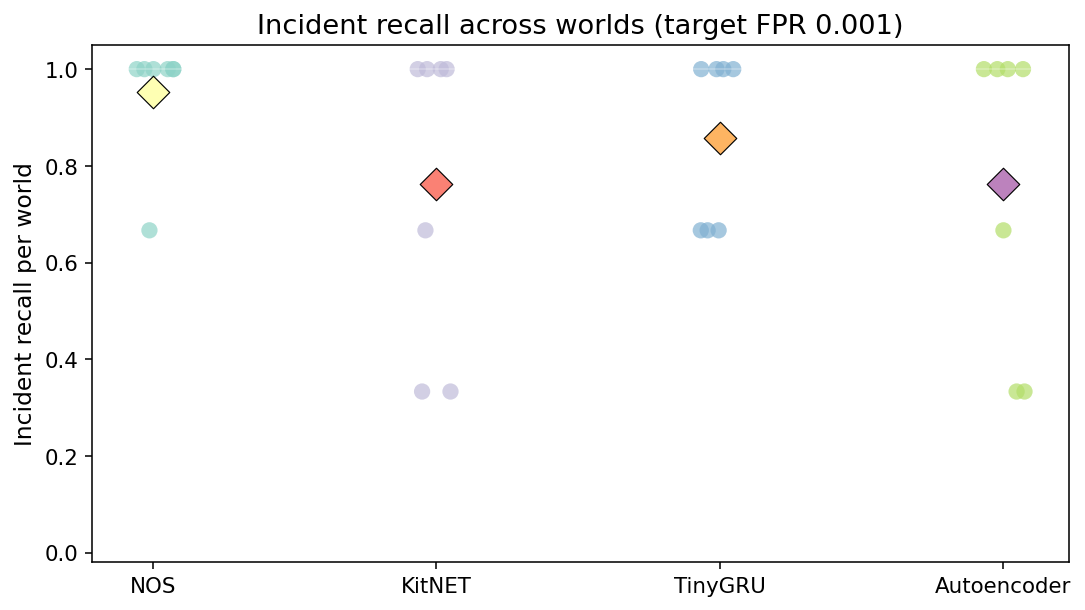}
\small (a) Incident recall per world ($\text{FPR}=0.1\%$).
\end{minipage}
\begin{minipage}{0.65\textwidth}
\centering
\includegraphics[width=\linewidth]{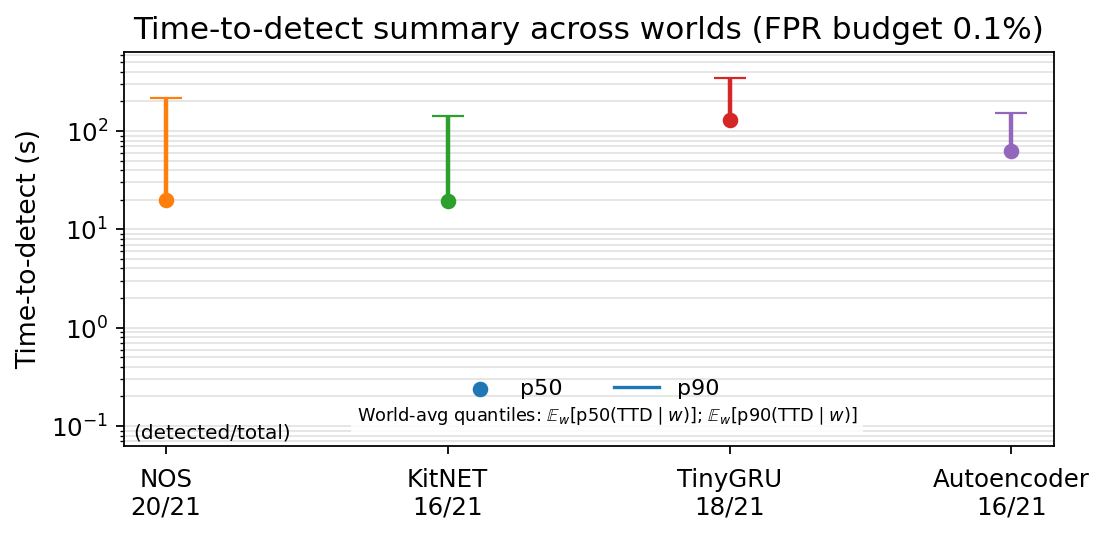}
\small (b) TTD distribution ($\text{FPR}=0.1\%$).
\end{minipage}
\begin{minipage}{0.65\textwidth}
\centering
\includegraphics[width=\linewidth]{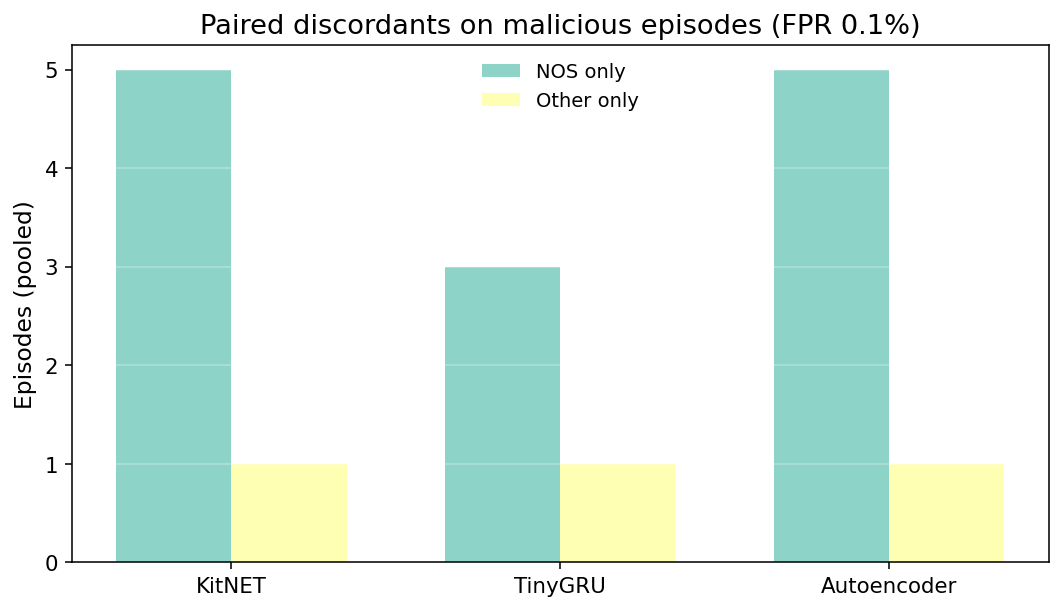}
\small (c) Discordant episode counts vs baselines ($\text{FPR}=0.1\%$).
\end{minipage}
\color{black}
\caption{Detection under the strict false-alarm budget (TTD is defined to the first alarm $a_{i,t}=1$, not to the first actionable trigger under $K$-of-$M$ persistence.). Panels (a) and (b) show per-world behaviour and latency dispersion, not only pooled totals. Panel (c) summarises episode-level discordants (NOS-Gate only vs baseline), pooled across worlds. }
\label{fig:det_main}
\end{figure*}

Figure~\ref{fig:det_main} summarises incident recall and time-to-detect (TTD) at $\text{FPR}=0.1\%$. \rev{Time-to-detect (TTD) is measured from the first malicious-labelled window of an episode to the first threshold-crossing alarm $a_{i,t}=1$ (i.e., the first window where $s_{i,t}\ge \vartheta_i$ on the attacked flow), and is reported separately from the persistence-based actionable signal used for WFQ gating. }This operating point is intentionally strict. Because $\vartheta_i$ is a high burn-in quantile and is then frozen, a detector must exceed a tail threshold on each flow. Under timing-controlled evasion (Table~\ref{tab:budgets}), the attacker can often reduce sharp per-window anomalies by spreading activity across windows while still meeting the throughput floor. This makes methods that depend on occasional high-amplitude windows easier to suppress below $\vartheta_i$, while methods that accumulate modest deviation energy across time are more likely to trigger without relying on a single extreme window.

At $\text{FPR}=0.1\%$, NOS-Gate misses 1 of 21 incidents (incident recall $0.952$). The strongest baseline in these runs, TinyGRU, misses 3 incidents (recall $0.857$). The remaining baselines miss 5 incidents (recall $0.762$).

Figure~\ref{fig:det_main}(a) shows per-world recall and indicates that the gain is not driven by a single outlier world. Figure~\ref{fig:det_main}(b) shows the TTD distribution over detected incidents. Several baselines can trigger quickly when they trigger, but under the strict budget they fail to trigger on more episodes. NOS-Gate trades some latency on the hardest cases for fewer missed incidents, which is the relevant trade-off when false alarms must remain rare.

Figure~\ref{fig:det_main}(c) provides episode-level discordant counts pooled across the same 21 episodes. NOS-Gate detects more episodes that each baseline misses than vice versa (KitNET: 5 vs 1; TinyGRU: 3 vs 1; autoencoder: 5 vs 1). The sample is modest, so we treat this as supporting evidence rather than a standalone statistical claim, but the direction is consistent across baselines.

\begin{figure*}[h!]
\centering
\begin{minipage}{0.6\textwidth}
\centering
\includegraphics[width=\linewidth]{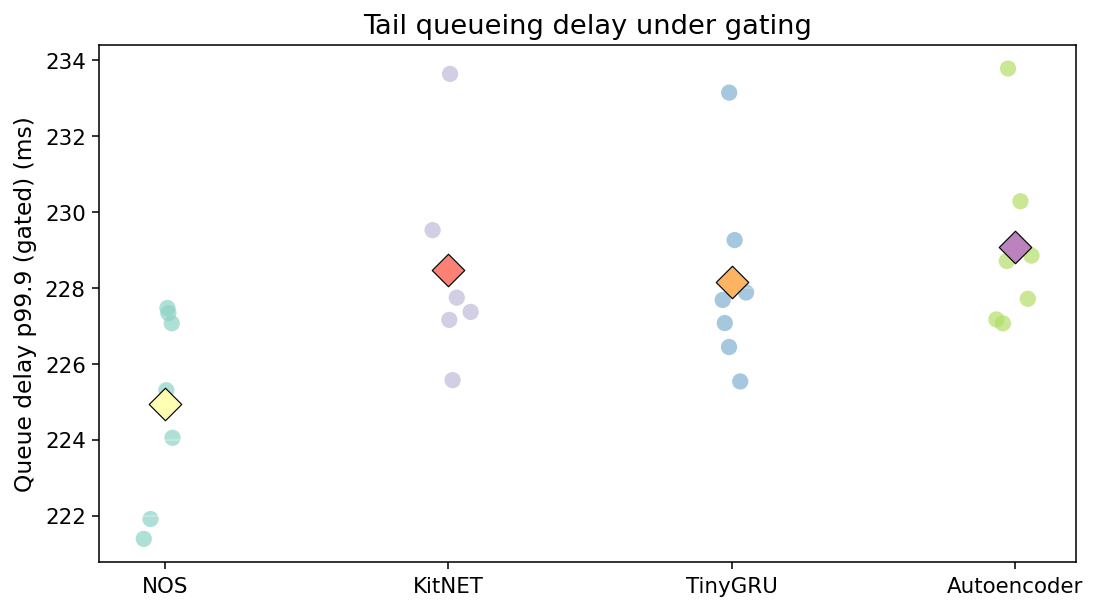}
\small (a) p99.9 queueing delay under gating.
\end{minipage}\hfill
\begin{minipage}{0.6\textwidth}
\centering
\includegraphics[width=\linewidth]{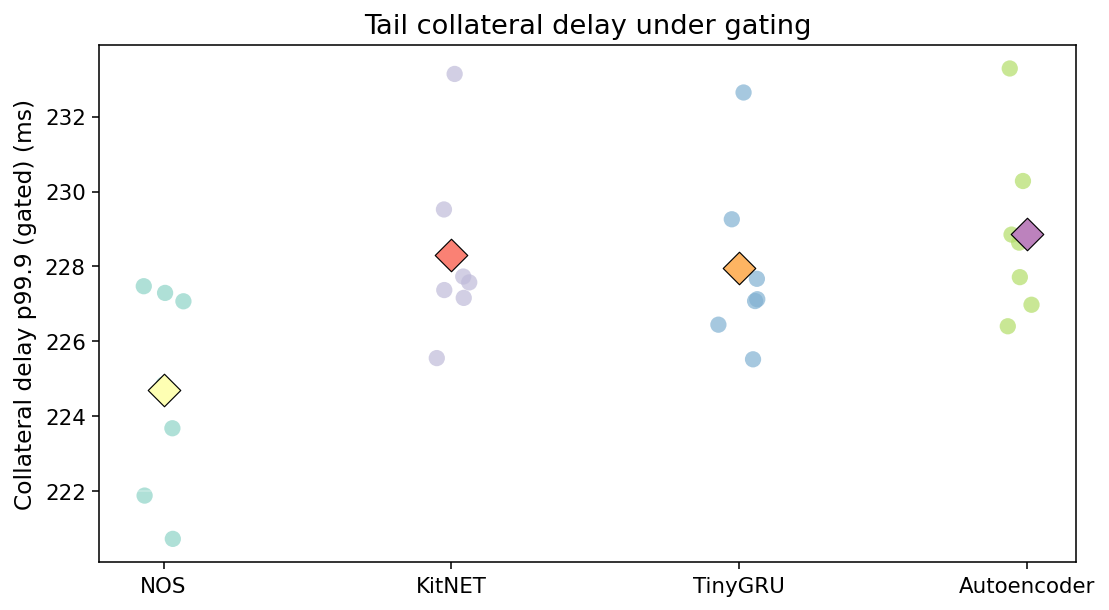}
\small (b) p99.9 collateral delay under gating.
\end{minipage}\hfill
\begin{minipage}{0.6\textwidth}
\centering
\includegraphics[width=\linewidth]{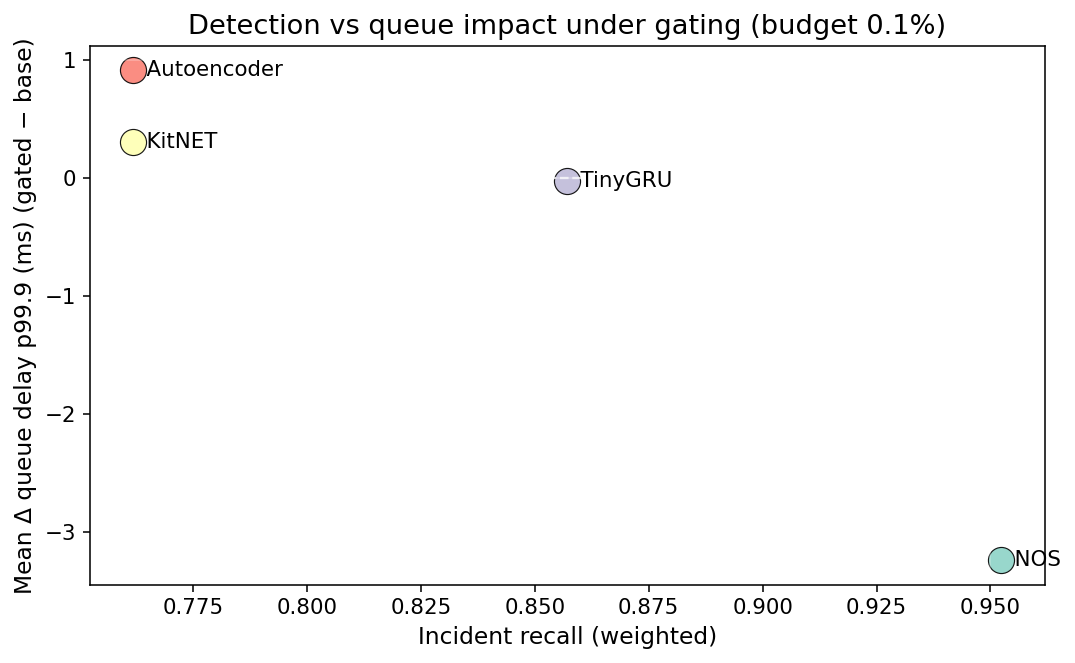}
\small (c) Incident recall vs $\Delta$p99.9 delay (gated minus base).
\end{minipage}
\caption{Action-level evaluation under the strict operating point ($\text{FPR}=0.1\%$). NOS-Gate is the only method that improves both queue-tail delay and collateral tail delay while maintaining the highest incident recall under the same label-free calibration protocol. Delays are reported in ms (raw logs are in $\mu$s).}
\label{fig:queue_main}
\end{figure*}
\subsection{Queue impact of WFQ gating}
Detection is evaluated here through its control effect. We therefore assess WFQ gating using packet-level replay, driven by each method’s actionable flags under the strict operating point ($\text{FPR}=0.1\%$). Queue logs are produced in microseconds and reported in milliseconds.

Figure~\ref{fig:queue_main} reports tail queueing delay and tail collateral delay under gating (p99.9). NOS-Gate is the only method that improves both metrics consistently across the 7 worlds. Relative to baseline replay without gating, NOS-driven gating reduces p99.9 queueing delay by about $3.24$ ms on average across worlds and reduces p99.9 collateral delay by about $3.16$ ms. In contrast, gating driven by KitNET and the autoencoder increases mean p99.9 queueing delay in these runs (about $+0.30$ ms and $+0.91$ ms, respectively), which is consistent with less targeted gating under the same strict alert budget.

Figure~\ref{fig:queue_main}(c) summarises the central deployment question: under a strict alert cap, does the detector sit in the favourable region of higher incident recall and lower tail delay. Under the shared stand-alone protocol, NOS-Gate occupies that region in these worlds.

\subsection{Efficiency on CPU}
Table~\ref{tab:eff} reports scoring cost in microseconds per window-row. NOS-Gate remains lightweight (mean $\approx 2.09\,\mu\text{s}$/row), comparable to the compact autoencoder and substantially faster than KitNET and TinyGRU in these runs. This matters because gateway cost scales with active flows and window cadence. A method that is expensive per window is harder to sustain without adding latency. \rev{All methods are timed under the same measurement protocol; we report per-row cost because gateway compute scales primarily with active flows and the fixed window cadence.}

\begin{table}[t]
\centering
\color{black}
\caption{Scoring cost on CPU (microseconds per window-row). Timings are wall-clock per $(\text{flow},\text{window})$ feature row on \textbf{Intel Core i7-1355U}, using \textbf{Windows 11} and \textbf{[Python 3.14.2]}; measurements were taken in \textbf{[single-thread]} mode. Reported time includes online normalisation and model update/scoring given precomputed window features; it excludes packet capture/parsing and window feature extraction.}
\label{tab:eff}
\begin{tabular}{lccc}
\toprule
Method & Mean ($\mu s$/row) & p90 ($\mu s$/row) & Max ($\mu s$/row) \\
\midrule
NOS-Gate    & 2.09 & 2.97 & 2.99 \\
Autoencoder & 2.03 & 2.78 & 2.80 \\
KitNET      & 11.68 & 16.72 & 16.80 \\
TinyGRU     & 31.10 & 46.57 & 49.24 \\
\bottomrule
\end{tabular}
\end{table}

\subsection{Summary}
Across these runs, NOS-Gate is the only method that meets the three constraints that define the gateway setting at once: (i) under a strict false-alarm budget it misses fewer incidents, (ii) its actionable flags drive a reversible WFQ action that reduces both queue-tail delay and collateral tail delay in replay, and (iii) it achieves this with a small, stable per-window compute cost. Under looser thresholds, baselines can appear closer. The strict stand-alone calibration exposes the failure mode that matters operationally: missed incidents and action-triggered latency when alerts must be rare.

\section{Conclusion}
Consumer gateways in the networks sit at the point where encrypted services meet heterogeneous IoT devices, but encryption leaves timing, size, and burst structure visible. Those side channels support metadata-based IDS, yet they also create a practical evasion surface: an adaptive adversary can reshape timing while still meeting an attack objective. This paper addressed a deployment-centred question: under a strict false-alarm cap and without labels, can a gateway detect quickly enough to act, using only metadata, and can that action reduce user-visible queue harm rather than add to it? We introduced \emph{NOS-Gate}, a streaming IDS that instantiates a lightweight two-state unit derived from Network-Optimised Spiking (NOS) dynamics per flow. The model is structurally faithful to NOS and changes only the interpretation of the exogenous drive, which becomes a scalar evidence signal derived from online normalised metadata features. The resulting state update provides controlled leak, bounded growth, and recovery-driven suppression, producing a stable score that is calibrated stand-alone by burn-in quantile thresholding and stabilised by a $K$-of-$M$ persistence rule. Crucially, NOS-Gate connects detection to an auditable mitigation path by temporarily reducing WFQ weights for persistently suspicious flows. To evaluate this action loop under timing-controlled evasion, we introduced an executable \emph{worlds} benchmark with explicit attacker budgets and packet-level WFQ replay. Under the strict operating point ($0.1\%$ achieved benign false-positive rate), NOS-Gate achieved the highest incident recall among the tested label-free baselines while also improving tail queue outcomes under gating, including reduced p99.9 queueing delay and reduced collateral tail delay. Its per-window scoring cost remained small and stable on CPU, which matches gateway scaling constraints.Finally, we do not claim payload-level coverage or that synthetic worlds capture every real deployment. Instead, we offer a falsifiable protocol, artefacts, and budgets that make both evasion and queue impact measurable. \rev{NOS-Gate operates per flow with a persistence rule, which is appropriate when malicious activity is sustained on a small number of flows. An adaptive adversary could attempt to dilute evidence by increasing flow churn (many short-lived flows) or by spreading activity across multiple endpoints so that no single flow accumulates enough alarms within the persistence window to trigger action. We do not evaluate such multi-flow strategies in this paper; addressing them may require lightweight aggregation above the five-tuple (for example, per-device or per-destination persistence) or coupling across flows sharing a scheduling domain, while preserving the same stand-alone calibration and auditable action constraints.} Future work should expand the benchmark with additional device processes, and explore how learned feature representations can be introduced without breaking the stand-alone calibration and action-audit requirements.


\bibliographystyle{IEEEtran}
\bibliography{refs}

\end{document}